\documentclass[aps,twocolumn,showpacs]{revtex4}
\usepackage{amsmath}
\usepackage{multirow}
\usepackage{epsfig}
\begin{document}

\title{Search for doubly-heavy dibaryons in the quark-delocalization color-screening model}

\author{Zhuocheng Xia$^1$}\email{xiazach003@outlook.com}
\author{Saijun Fan$^1$}\email{969985589@qq.com}
\author{Xinmei Zhu$^2$}\email{zxm_yz@126.com}
\author{Hongxia Huang$^1$}\email{hxhuang@njnu.edu.cn(Corresponding author)}
\author{Jialun Ping$^1$}\email{jlping@njnu.edu.cn(Corresponding author)}

\affiliation{$^1$Department of Physics and Jiangsu Key Laboratory for Numerical
Simulation of Large Scale Complex Systems, Nanjing Normal University, Nanjing 210023, P. R. China}
\affiliation{$^2$Department of Physics, Yangzhou University, Yangzhou 225009, P. R. China}

\begin{abstract}
We perform a systemical investigation of the low-lying doubly-heavy dibaryon systems with strange $S=0$, isospin $I=0$, $1$, $2$ and the angular momentum $J=0$, $1$, $2$, $3$ in the quark delocalization color screening model. We find the effect of channel-coupling cannot be neglected in the study of the multi-quark systems. Due to the heavy flavor symmetry, the results of the doubly-charm and doubly-bottom dibaryon systems are similar with each other. Both of them have three bound states, the quantum numbers of which are $IJ=00$, $IJ=02$ and $IJ=13$, respectively. The energies are $4554$ MeV, $4741$ MeV, and $4969$ MeV respectively for the doubly-charm systems and $11219$ MeV, $11416$ MeV, and $11633$ MeV respectively for the doubly-bottom dibaryon systems. Besides, six resonance states are obtained, which are $IJ=00$ $N\Xi_{cc}$ and $N\Xi_{bb}$ with resonance mass of $4716$ MeV and $11411$ MeV respectively, $IJ=11$ $N\Xi^*_{cc}$ and $N\Xi^*_{bb}$ with resonance mass of $4757$ MeV and $11432$ MeV respectively,
and $IJ=12$ $\Sigma_{c}\Sigma^*_{c}$ and $\Sigma_{b}\Sigma^*_{b}$ with resonance mass of $4949$ MeV and $11626$ MeV respectively. All these heavy dibaryons are worth searching for on experiments, although it will be a challenging work.
\end{abstract}


\maketitle

\setcounter{totalnumber}{5}

\section{\label{sec:introduction}Introduction}
A worldwide theoretical and experimental effort to search for dibaryons has lasted a long time. Although the research on the dibaryon has experienced several ups and downs in its history, dibaryons have received renewed interest in recent years. The well-known dibaryon resonance $d^{*}$ was repeatedly observed by the WASA detector at COSY~\cite{ABC1,ABC2,ABC3,ABC4,ABC5}, and extensively investigated within various theoretical approaches~\cite{Dyson,Goldman,faddeev,Bashkanov,Ping_NN,Huang_NN,DongYB_NN,ChenHX_NN}.
Another dibaryon $N\Omega$ was proposed as a narrow resonance in a relativistic quark model~\cite{PRL59}, and was investigated by other quark models~\cite{Oka,PRC51,PangHR,ChenM,Huang2,LiQB}, as well as the lattice QCD~\cite{HAL1,HAL2}. The progress of the $N\Omega$ searches by the STAR experiment at the Relativistic Heavy-Ion Collider (RHIC) favored the existence of $N\Omega$~\cite{RHIC}. Besides, the study of the strong interaction among hadrons at the LHC by ALICE Collaboration also supported the possibility of forming the $N\Omega$ state~\cite{ALICE}.

In the past decade, many near-threshold charmonium-like states called $XYZ$ particles were observed, triggering lots of studies on the molecule-like bound states containing heavy quark hadrons. Such studies will give further information on the hadron-hadron interactions. In the heavy-quark sector, the large masses of the heavy quarks reduce the kinetic energy of the system, which makes them easier to form bound states. Therefore, extending the dibaryon research to the heavy quark sector is spontaneous.
The $N\Lambda_{c}$ system and the $H$-like dibaryon state $\Lambda_{c}\Lambda_{c}$ have been studied on both hadron level~\cite{Liu1,Liu2} and quark level~\cite{Huang_NL,Huang_LL}. The possibility of existing deuteron-like dibaryons with heavy quarks, such as
$N\Sigma_{c}$, $N\Xi^{'}_{c}$, $N\Xi_{cc}$, $\Xi\Xi_{cc}$ and so on, were investigated by several realistic phenomenological
nucleon-nucleon interaction models~\cite{Fromel,Julia}. In Ref.~\cite{MengL}, the deuteron-like states composed of two doubly charmed baryons $\Xi_{cc}\Xi_{cc}$ and $\Xi_{cc}\overline{\Xi}_{cc}$ were systematically studied within the one-boson-exchange model. Besides, the possible $N\Omega$-like dibaryons $N\Omega_{ccc}$ and $N\Omega_{bbb}$ were investigated within both the chiral quark model and quark delocalization color screening model~\cite{Huang_NO}.
Recently, Junnarkar and Mathur reported the first lattice QCD study of deuteron-like dibaryons with heavy quark flavors~\cite{Lattice_De}, and suggested that the dibaryons $\Omega_{c}\Omega_{cc} (sscscc)$, $\Omega_{b}\Omega_{bb} (ssbsbb)$, and $\Omega_{ccb}\Omega_{cbb} (ccbcbb)$ were stable under strong and electromagnetic interactions. They also found that the binding of these dibaryons became stronger as they became heavier in mass. However, the distinct conclusion was claimed in the work of Ref.~\cite{Richard}, where the authors explored the possibility of very heavy dibaryons with three charm quarks and three beauty quarks ($bbbccc$) in potential models, and concluded that there was no evidence for any stable state in such very heavy flavored six-quark system. Additionally, the existence of fully heavy dibaryons were also investigated in the constituent quark model~\cite{Huang_OO}.

Quantum chromodynamics (QCD) is widely accepted as a fundamental theory to study strong interaction. However, for hadron-hadron interactions and exotic quark states, it is difficult to use QCD directly to study low-energy hadronic interaction because of the nonperturbative complication. Therefore, it has triggered kinds of QCD-inspired models, which study the muliquark systems from the physical perspective. The quark delocalization color screening model (QDCSM) is one of the representations of the constituent quark models, which was developed in the 1990s, aiming to explain the similarities between nuclear and molecular force~\cite{QDCSM0}. The model modifies the Hamiltonian by introducing the shielding effect of color screen and expands Hilbert space by considering the quark delocalization between two quark clusters. This model has been well applied to describe the properties of the deuteron~\cite{QDCSM1}, study the $NN$ and $YN$ interactions, and investigate the dibaryon candidates~\cite{QDCSM2,QDCSM3,QDCSM4,QDCSM5}. It has also been extended to observe the dibaryon states with heavy quarks, such as the $N\Lambda_{c}$ and $N\Lambda_{b}$ systems~\cite{Huang_NL}, and the possible $H-$like dibaryon states $\Lambda_{c}\Lambda_{c}$ and $\Lambda_{b}\Lambda_{b}$~\cite{Huang_LL}, the $N\Omega$-like dibaryons $N\Omega_{ccc}$ and $N\Omega_{bbb}$~\cite{Huang_NO}, the fully heavy dibaryons~\cite{Huang_OO}, and so on. It is interesting to search for more dibaryons in heavy quark sector within the frame of QDCSM.

In this work, we further study the doubly-heavy dibaryons systematically in the QDCSM. Extension of the study to the bottom case is also interesting and is performed too. Through our calculation, we can look for possible doubly-heavy dibaryons, which will provide more information for the experimental work. The structure of this paper is as follows. After the introduction, we have a simple description of the QDCSM in section II. Section III is the numerical results and discussions. The summery is given in the last section.

\section{Quark delocalization color screening model}
The quark delocalization and color screening model (QDCSM) has been described in detail in Refs.~\cite{QDCSM0,QDCSM1,QDCSM2,QDCSM3}. Here,we just present the Hamiltonian of the model.
\begin{widetext}
\begin{eqnarray}
H &=& \sum_{i=1}^6 \left(m_i+\frac{p_i^2}{2m_i}\right) -T_c
+\sum_{i<j} \left[ V^{G}(r_{ij})+V^{\chi}(r_{ij})+V^{C}(r_{ij})
\right],
 \nonumber \\
V^{G}(r_{ij})&=& \frac{1}{4}\alpha_{s} {\mathbf \lambda}_i \cdot
{\mathbf \lambda}_j
\left[\frac{1}{r_{ij}}-\frac{\pi}{2}\left(\frac{1}{m_{i}^{2}}+\frac{1}{m_{j}^{2}}+\frac{4{\mathbf
\sigma}_i\cdot {\mathbf\sigma}_j}{3m_{i}m_{j}}
 \right)
\delta(r_{ij})-\frac{3}{4m_{i}m_{j}r^3_{ij}}S_{ij}\right],
\nonumber \\
V^{\chi}(r_{ij})&=& \frac{1}{3}\alpha_{ch}
\frac{\Lambda^2}{\Lambda^2-m_{\chi}^2}m_\chi \left\{ \left[
Y(m_\chi r_{ij})- \frac{\Lambda^3}{m_{\chi}^3}Y(\Lambda r_{ij})
\right]
{\mathbf \sigma}_i \cdot{\mathbf \sigma}_j \right.\nonumber \\
&& \left. +\left[ H(m_\chi r_{ij})-\frac{\Lambda^3}{m_\chi^3}
H(\Lambda r_{ij})\right] S_{ij} \right\} {\mathbf F}_i \cdot
{\mathbf F}_j, ~~~\chi=\pi,K,\eta \\
V^{C}(r_{ij})&=& -a_c {\mathbf \lambda}_i \cdot {\mathbf
\lambda}_j [f(r_{ij})+V_0], \nonumber
\\
 f(r_{ij}) & = &  \left\{ \begin{array}{ll}
 r_{ij}^2 &
 \qquad \mbox{if }i,j\mbox{ occur in the same baryon orbit} \\
  \frac{1 - e^{-\mu_{ij} r_{ij}^2} }{\mu_{ij}} & \qquad
 \mbox{if }i,j\mbox{ occur in different baryon orbits} \\
 \end{array} \right.
\nonumber \\
S_{ij} & = &  \frac{{\mathbf (\sigma}_i \cdot {\mathbf r}_{ij})
({\mathbf \sigma}_j \cdot {\mathbf
r}_{ij})}{r_{ij}^2}-\frac{1}{3}~{\mathbf \sigma}_i \cdot {\mathbf
\sigma}_j. \nonumber
\end{eqnarray}
\end{widetext}
where $S_{ij}$ is the quark tensor operator; $Y(\chi)$ and $H(\chi)$ are standard Yukawa functions~\cite{Salamanca}; $T_{CM}$ is the kinetic energy of the center; and $\alpha_s$ is the quark-gluon coupling constant. In order to cover the wide energy range from light, strange to heavy quarks, one introduces an effective scale-dependent quark-gluon coupling constant $\alpha_s(\mu)$~\cite{JPG31}.
\begin{equation}
\alpha_s(\mu)=\frac{\alpha_{0}}{\ln\left(\frac{\mu^{2}+\mu_0^{2}}{\Lambda_0^{2}}\right)}.
\label{alpha-s}
\end{equation}
where $\mu$ is the reduced mass of the interacting quark pair.
The coupling constant $g_{ch}$ for scalar chiral field is determined from the $NN\pi$ coupling constant through
\begin{equation}
\frac{g_{ch}^{2}}{4\pi }=\left( \frac{3}{5}\right) ^{2}{\frac{g_{\pi NN}^{2}%
}{{4\pi }}}{\frac{m_{u,d}^{2}}{m_{N}^{2}}}\label{gch}
\end{equation}
All other symbols have their usual meanings in the above expressions. Here, all the parameters related to the light quarks are from our previous study of strange dibaryons~\cite{QDCSM3}. Other parameters related to the heavy quarks are adjusted by fitting the masses of the charmed and bottom baryons. All parameter values are listed in Table~\ref{parameters}. Table~\ref{mass} lists the corresponding masses of the charmed and bottom baryons.
\begin{table}[ht]
\caption{Model parameters: $m_{\pi}=0.7~{\rm fm}^{-1}$, $m_{K}=2.51~{\rm fm}^{-1}$,
$m_{\eta}=2.77~{\rm fm}^{-1}$, $\Lambda_{\pi}=4.2~{\rm fm}^{-1}$,
$\Lambda_{K}=5.2~{\rm fm}^{-1}$, $\Lambda_{\eta}=5.2~{\rm
fm}^{-1}$, $\alpha_{ch}=0.027$.}
\begin{tabular}{lccccc} \hline
& ~~~~~~$b$~~~~ & ~~~~$m_{u,d}$~~~~ & ~~~~$m_{s}$~~~~ & ~~~~$m_{c}$~~~~ & ~~~~$m_{b}$~~~~   \\
& (fm) & (MeV) & (MeV) & (MeV) & (MeV)   \\ \hline\noalign{\smallskip}
QDCSM & 0.6  & 313  &  539 & 1732 &    5070   \\ \hline\noalign{\smallskip}
 & $ a_c$ & $V_{0}$ &  $\Lambda_{0}$ & $u_{0}$ &  $\alpha_{0}$   \\
 & (MeV\,fm$^{-2}$) & (fm$^{2}$) &  (fm$^{-1}$) & (MeV) &   \\\hline\noalign{\smallskip}
QDCSM & 18.5283 & -0.3333  &  1.7225 &   445.8512 &  0.7089  \\
\hline
\end{tabular}
\label{parameters}
\end{table}

\begin{table}[ht]
\caption{\label{mass}The masses (in MeV) of the baryons used in this work. Experimental values are taken from the Particle Data Group(PDG)~\cite{PDG}}
\begin{tabular}{lcccccccc}
\hline \hline
               & ~~$N$~~              & ~~$\Delta$~~      & ~~$\Lambda_{c}$~~   & ~~$\Sigma_{c}$~~
               & ~~$\Sigma^*_{c}$~~   & ~~$\Xi_{cc}$~~    & ~~$\Xi^*_{cc}$~~  \\ \hline
QDCSM          & 939          & 1232       & 2286    & 2462
               & 2492         & 3794       & 3823         \\
 Exp.          & 939  &1233 &2286 &2455 &2520 &-- &3621  \\
               & ~~$\Lambda_{b}$~~   & ~~$\Sigma_{b}$~~  & ~~$\Sigma^*_{b}$~~   & ~~$\Xi_{bb}$~~   & ~~$\Xi^*_{bb}$~~\\
QDCSM          & 5619          & 5809       & 5818    & 10485      & 10494    \\
 Exp.          & 5619          & 5811       & 5832    & --         & --  \\  \hline\hline
\end{tabular}
\end{table}

The quark delocalization in QDCSM is realized by specifying the
single particle orbital wave function of QDCSM as a linear
combination of left and right Gaussians, the single particle
orbital wave functions used in the ordinary quark cluster model,
\begin{eqnarray}
\psi_{\alpha}(\mathbf{s}_i ,\epsilon) & = & \left(
\phi_{\alpha}(\mathbf{s}_i)
+ \epsilon \phi_{\alpha}(-\mathbf{s}_i)\right) /N(\epsilon), \nonumber \\
\psi_{\beta}(-\mathbf{s}_i ,\epsilon) & = &
\left(\phi_{\beta}(-\mathbf{s}_i)
+ \epsilon \phi_{\beta}(\mathbf{s}_i)\right) /N(\epsilon), \nonumber \\
N(\epsilon) & = & \sqrt{1+\epsilon^2+2\epsilon e^{-s_i^2/4b^2}}. \label{1q} \\
\phi_{\alpha}(\mathbf{s}_i) & = & \left( \frac{1}{\pi b^2}
\right)^{3/4}
   e^{-\frac{1}{2b^2} (\mathbf{r}_{\alpha} - \mathbf{s}_i/2)^2} \nonumber \\
\phi_{\beta}(-\mathbf{s}_i) & = & \left( \frac{1}{\pi b^2}
\right)^{3/4}
   e^{-\frac{1}{2b^2} (\mathbf{r}_{\beta} + \mathbf{s}_i/2)^2}. \nonumber
\end{eqnarray}
Here $\mathbf{s}_i$, $i=1,2,...,n$ are the generating coordinates,
which are introduced to expand the relative motion
wavefunction. The delocalization parameter $\epsilon(\mathbf{s}_i)$ is
determined by the dynamics of the multi-quark system. In this way, the system can choose its most
favorable configuration through its own dynamics in a larger
Hilbert space. It has been used to explain the cross-over transition between hadron phase and
quark-gluon plasma phase~\cite{Xu}.

\section{The results and discussions}
In this work, we perform a systematical investigation to the low-lying doubly-heavy dibaryon systems with strange $S=0$, isospin $I=0$, $1$, $2$ and the angular momentum $J=0$, $1$, $2$, $3$. Since the attractive potential is necessary for forming a bound state or resonance, we first calculate the effective potential between two baryons, which is defined as $V(S)=E(S)-E(\infty)$, where $E(S)$ is the diagonal matrix element of the Hamiltonian of the system in the generating coordinate. Here, we first show the results of the doubly-charm dibaryon systems, and the one of the doubly-bottom dibaryon systems will be shown at the end of this section. The effective potentials of all channels with different quantum numbers shown in Figs. 1, 2 and 3, respectively.

From Figs. 1, 2 and 3, we can see that the effective potentials of several channels are purely repulsive, which are $\Lambda_{c}\Lambda_{c}$ with $IJ=00$, $N\Xi^*_{cc}$ with $IJ=01$, $N\Xi_{cc}$ and $\Lambda_{c}\Sigma_{c}$ with $IJ=10$, $\Lambda_{c}\Sigma^*_{c}$, $N\Xi_{cc}$ and $\Lambda_{c}\Sigma_{c}$ with $IJ=11$, $\Lambda_{c}\Sigma^*_{c}$ and $N\Xi^*_{cc}$ with $IJ=12$, $\Delta\Xi^*_{cc}$ with $IJ=21$, $\Delta\Xi_{cc}$ and $\Sigma^*_{c}\Sigma^*_{c}$ with $IJ=22$, and $\Delta\Xi^*_{cc}$ with $IJ=23$. So it is difficult for these channels to form any bound state. Conversely, the following channels have a deep effective attraction, which is larger than $-100$ MeV. They are $\Sigma^*_{c}\Sigma^*_{c}$ and $\Sigma_{c}\Sigma_{c}$ with $IJ=00$, $\Sigma_{c}\Sigma^*_{c}$ with $IJ=01$, $\Sigma^*_{c}\Sigma^*_{c}$ with $IJ=02$, $\Delta\Xi^*_{cc}$ with $IJ=10$, $\Delta\Xi^*_{cc}$ and $\Delta\Xi_{cc}$ with $IJ=11$, and $\Delta\Xi^*_{cc}$ and $\Delta\Xi_{cc}$ with $IJ=12$. Such deep attraction will make these channels more likely to form bound states or resonance states. For other channels, the effective potentials are attractive too. Although the attraction is not very deep, we still need to verdict the existence of bound states or resonance states for these channels.

\begin{figure*}[htb]
\centering
\includegraphics[totalheight=8.0cm]{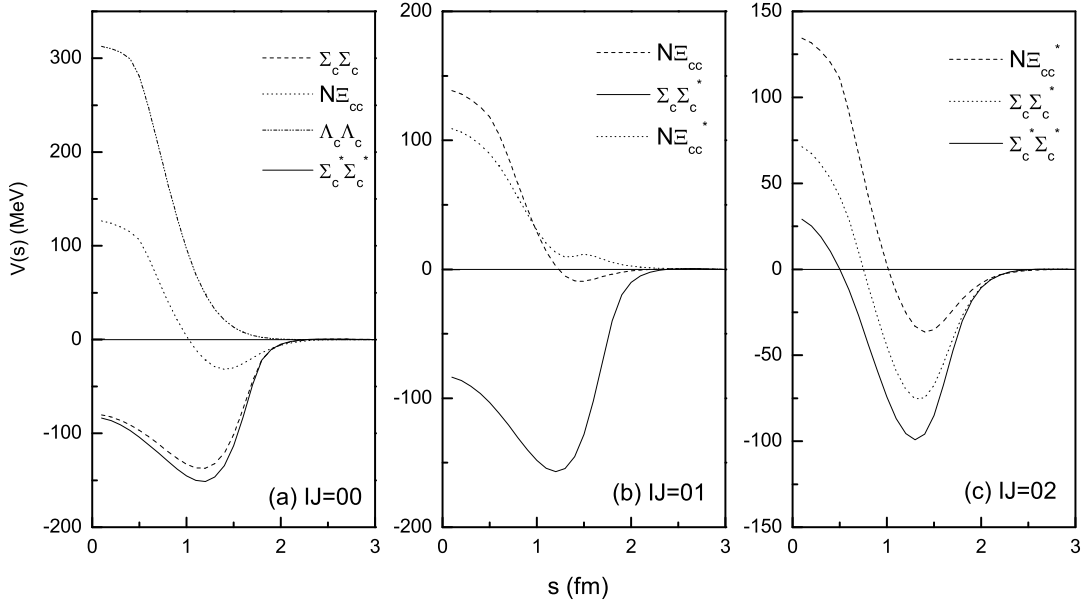}\\
\caption{The effective potentials of different channels of the doubly-charm dibaryon systems with $I=0$.} \label{0}
\end{figure*}

\begin{figure*}[htb]
\centering
\includegraphics[totalheight=6.5cm]{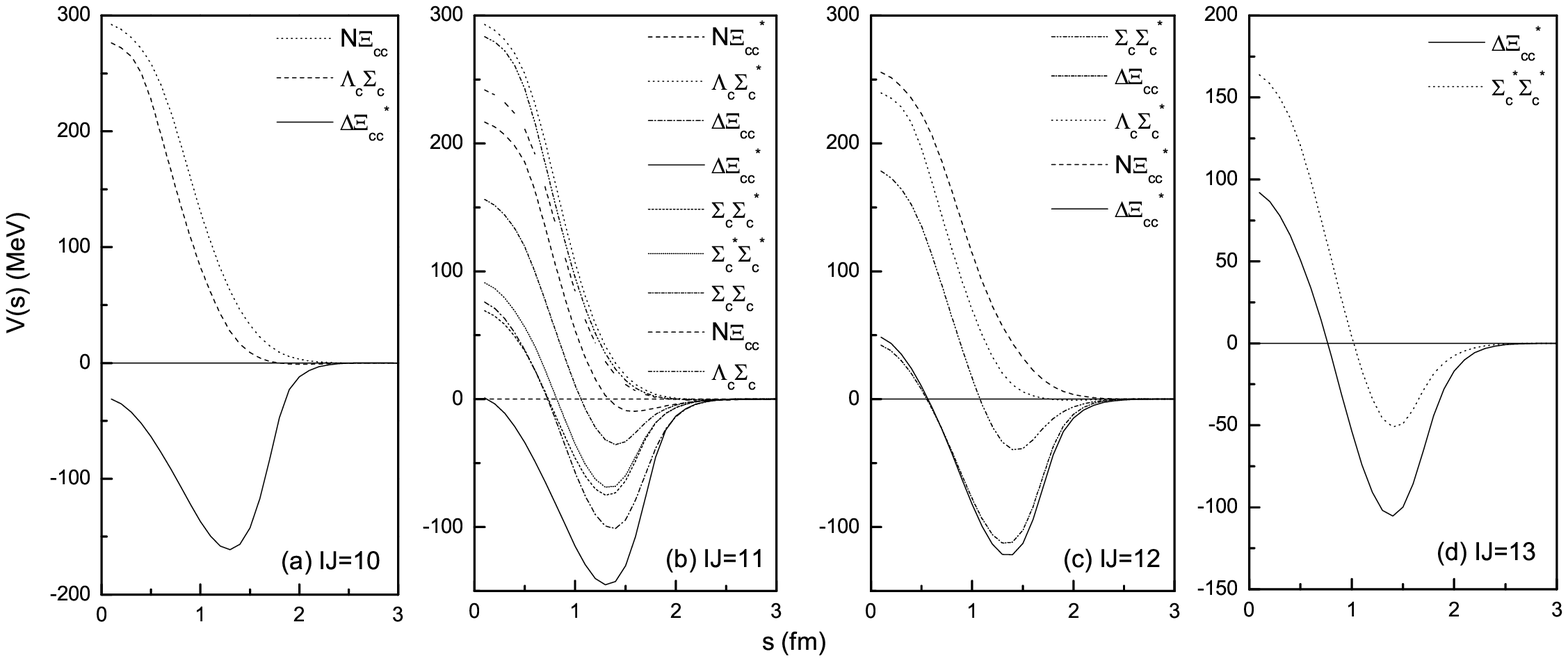}\\
\caption{The effective potentials of different channels of the doubly-charm dibaryon systems with $I=1$.} \label{1}
\end{figure*}

\begin{figure*}[htb]
\centering
\includegraphics[totalheight=6.5cm]{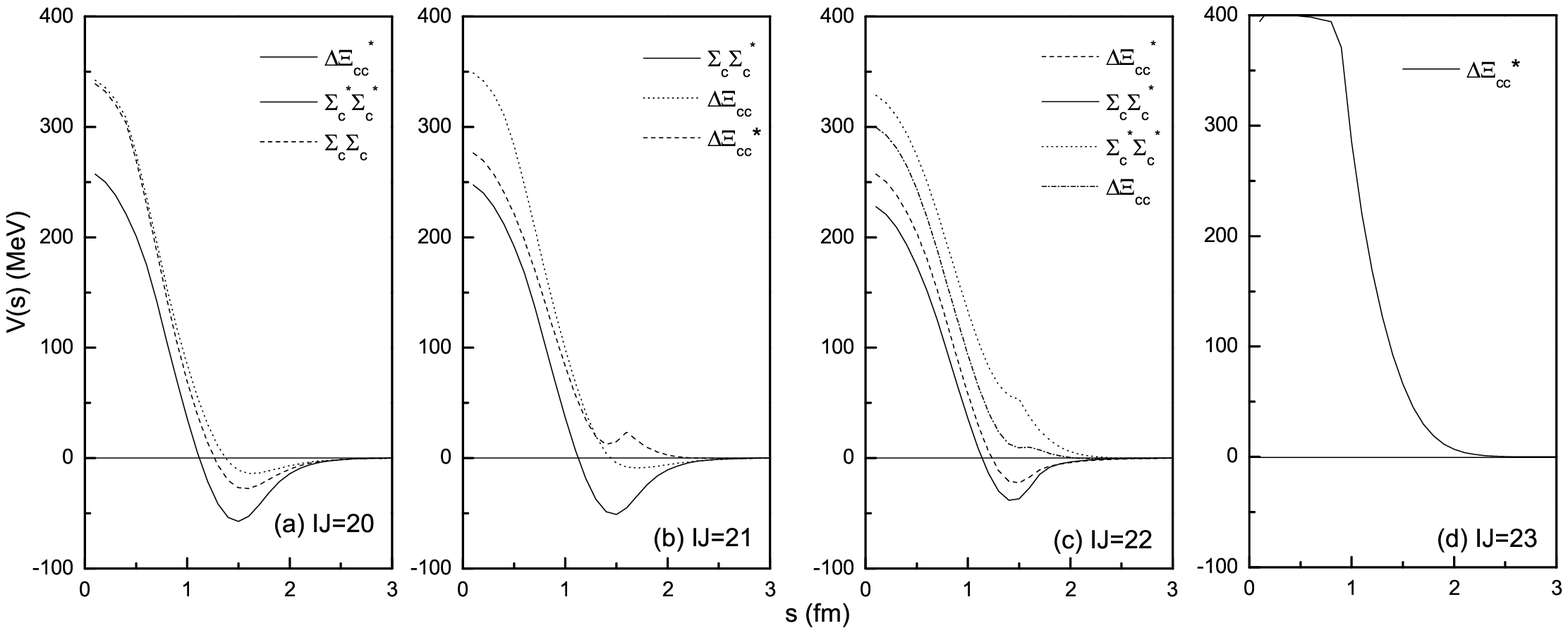}\\
\caption{The effective potentials of different channels of the doubly-charm dibaryon systems with $I=2$.} \label{2}
\end{figure*}

In order to see whether or not there is any bound state, a dynamic calculation
based on the resonating group method (RGM)~\cite{RGM} has been performed. By expanding
the relative motion wave function between two clusters in the RGM equation by gaussians, the
integro-differential equation of RGM can be reduced to an algebraic equation, which is the generalized eigen-equation. Then the energy of the system can be obtained by solving the eigen-equation. Besides, to keep the matrix dimension manageably small, the baryon-baryon separation is taken to be less
than $6$ fm in the calculation. The energy of each channel, as well as the one with channel-coupling calculation are listed in Table~\ref{result1}, where the first column is the quantum number of the system; the second column is the corresponding state of every channel; the third column $E_{th}$ denotes the theoretical threshold of each state; the fourth column $E_{sc}$ represents the energy of every single channel; the fifth column $B_{sc}$ stands for the binding energy of every single channel, which is $B_{sc}=E_{sc}-E_{th}$; the sixth column $E_{cc}$ denotes the lowest energy of the system by channel-coupling calculation; and the last column $B_{cc}$ represents the binding energy with all channels coupling, which is $B_{cc}=E_{cc}-E_{th}$. Here, we should notice that the positive value of the binding energy $B_{sc}$ or $B_{cc}$ means that the state is unbound, so we label as 'ub' in Table~\ref{result1}. In addition, the channel with the lowest threshold of each system is bolded in order to distinguish clearly in the table.

\begin{table*}[ht]
\caption{The energy of the doubly-charm dibaryon systems.}   \label{result1}
\centering
\begin{tabular}{ccccccc}
\hline \hline
  ~~ $IJ$ &  ~~ $Channels$ &  ~~$E_{th}$~(MeV) &  ~~$E_{sc}$~(MeV) &  ~~$B_{sc}$~(MeV) & ~~$E_{cc}$~(MeV) &  ~~$B_{cc}$~(MeV)
  \\ \hline
  ~~\multirow{4}{*}{00} &  ~~$\Sigma_{c}\Sigma_{c}$ &    ~~4925 &    ~~4850 &   ~~-75 &  ~~\multirow{4}{*}{4554} &  ~~\multirow{4}{*}{-19}    \\ \cline{3-5}
  ~~  &  ~~$N\Xi_{cc}$ &                           ~~4733 &   ~~4736 &    ~~ub  &     ~~ &      \\ \cline{3-5}
  ~~  &  ~~$\mathbf{\Lambda_{c}\Lambda_{c}}$ &   ~~4573 &   ~~4580 &    ~~ub  &     ~~ &      \\ \cline{3-5}
  ~~  &  ~~$\Sigma^*_{c}\Sigma^*_{c}$ &            ~~4983 &   ~~4896 &    ~~-87 &     ~~ &      \\ \hline
  ~~\multirow{3}{*}{01} &  ~~$\mathbf{N\Xi_{cc}}$ &    ~~4733&     ~~4739 &   ~~ub &   ~~\multirow{3}{*}{4733} &   ~~\multirow{3}{*}{ub}   \\ \cline{3-5}
  ~~  &  ~~$\Sigma_{c}\Sigma^*_{c}$ &              ~~4954 &   ~~4860 &    ~~-94 &     ~~ &     \\ \cline{3-5}
  ~~  &  ~~$N\Xi^*_{cc}$ &                         ~~4762 &   ~~4770 &    ~~ub  &     ~~ &     \\ \hline
  ~~\multirow{3}{*}{02}&  ~~$\mathbf{N\Xi^*_{cc}}$ &  ~~4762 &    ~~4779 &   ~~ub &   ~~\multirow{3}{*}{4741} &   ~~\multirow{3}{*}{-21}  \\\cline{3-5}
  ~~  &  ~~$\Sigma_{c}\Sigma^*_{c}$ &              ~~4954 &   ~~4965 &    ~~ub  &     ~~ &     \\ \cline{3-5}
  ~~  &  ~~$\Sigma^*_{c}\Sigma^*_{c}$ &            ~~4983 &   ~~4993 &    ~~ub  &     ~~ &     \\ \hline
  ~~\multirow{3}{*}{10} &  ~~$\mathbf{N\Xi_{cc}}$ &    ~~4733 &    ~~4740 &   ~~ub &   ~~\multirow{3}{*}{4740} &   ~~\multirow{3}{*}{ub} \\ \cline{3-5}
  ~~  &  ~~$\Lambda_{c}\Sigma_{c}$ &               ~~4749 &   ~~4756 &    ~~ub  &     ~~ &     \\ \cline{3-5}
  ~~  &  ~~$\Delta\Xi^*_{cc}$ &                    ~~5055 &   ~~4956 &    ~~-99  &    ~~ &     \\ \hline
  ~~\multirow{9}{*}{11} &  ~~$N\Xi^*_{cc}$ &             ~~4762 &    ~~4768 &   ~~ub &   ~~\multirow{9}{*}{4739} &  ~~\multirow{9}{*}{ub} \\ \cline{3-5}
  ~~  &  ~~$\Lambda_{c}\Sigma^*_{c}$ &             ~~4778 &   ~~4785 &    ~~ub   &    ~~ &     \\ \cline{3-5}
  ~~  &  ~~$\Delta\Xi_{cc}$ &                      ~~5026 &   ~~4986 &    ~~-40  &    ~~ &     \\ \cline{3-5}
  ~~  &  ~~$\Delta\Xi^*_{cc}$ &                    ~~5055 &   ~~4971 &    ~~-84  &    ~~ &     \\ \cline{3-5}
  ~~  &  ~~$\Sigma_{c}\Sigma^*_{c}$ &              ~~4954 &   ~~4937 &    ~~-17  &    ~~ &     \\ \cline{3-5}
  ~~  &  ~~$\Sigma^*_{c}\Sigma^*_{c}$ &            ~~4983 &   ~~4971 &    ~~-12  &    ~~ &     \\ \cline{3-5}
  ~~  &  ~~$\Sigma_{c}\Sigma_{c}$ &                ~~4925 &   ~~4928 &    ~~ub   &    ~~ &     \\ \cline{3-5}
  ~~  &  ~~$\mathbf{N\Xi_{cc}}$ &                ~~4733 &   ~~4740 &    ~~ub   &    ~~ &     \\ \cline{3-5}
  ~~  &  ~~$\Lambda_{c}\Sigma_{c}$ &               ~~4749 &   ~~4756 &    ~~ub   &    ~~ &     \\ \hline
  ~~\multirow{5}{*}{12} &  ~~$\Sigma_{c}\Sigma^*_{c}$ &  ~~4954 &    ~~4956 &   ~~ub &   ~~\multirow{5}{*}{4769} &  ~~\multirow{5}{*}{ub} \\ \cline{3-5}
  ~~  &  ~~$\Delta\Xi_{cc}$ &                   ~~5026 &   ~~4975 &    ~~-51  &    ~~ &     \\ \cline{3-5}
  ~~  &  ~~$\Lambda_{c}\Sigma^*_{c}$ &             ~~4778 &   ~~4785 &    ~~ub   &    ~~ &     \\ \cline{3-5}
  ~~  &  ~~$\mathbf{N\Xi^*_{cc}}$ &              ~~4762 &   ~~4770 &    ~~ub   &    ~~ &     \\ \cline{3-5}
  ~~  &  ~~$\Delta\Xi^*_{cc}$ &                    ~~5055 &   ~~5065 &    ~~ub   &    ~~ &     \\ \hline
  ~~\multirow{2}{*}{13}  &  ~~$\Delta\Xi^*_{cc}$ &       ~~5055 &    ~~5065 &   ~~ub &   ~~\multirow{2}{*}{4969} &  ~~\multirow{2}{*}{-14} \\\cline{3-5}
  ~~  &  ~~$\mathbf{\Sigma^{*}_{c}\Sigma^{*}_{c}}$ & ~~4983 &   ~~4980 &    ~~-3   &    ~~ &     \\ \hline
  ~~\multirow{3}{*}{20} &  ~~$\Delta\Xi^*_{cc}$ &        ~~5055 &    ~~5047 &   ~~-8 &   ~~\multirow{3}{*}{4929} &   ~~\multirow{3}{*}{ub} \\ \cline{3-5}
  ~~  &  ~~$\Sigma^*_{c}\Sigma^*_{c}$ &            ~~4983 &   ~~4986 &    ~~ub   &    ~~ &     \\ \cline{3-5}
  ~~  &  ~~$\mathbf{\Sigma_{c}\Sigma_{c}}$ &     ~~4925 &   ~~4930 &    ~~ub   &    ~~ &     \\ \hline
  ~~\multirow{3}{*}{21} &  ~~$\mathbf{\Sigma_{c}\Sigma^*_{c}}$ &   ~~4954 &   ~~4959 &   ~~ub & ~~\multirow{3}{*}{4958} &   ~~\multirow{3}{*}{ub} \\ \cline{3-5}
  ~~  &  ~~$\Delta\Xi_{cc}$ &                      ~~5026 &   ~~5033 &    ~~ub   &    ~~ &     \\ \cline{3-5}
  ~~  &  ~~$\Delta\Xi^*_{cc}$ &                    ~~5055 &   ~~5069 &    ~~ub   &    ~~ &     \\ \hline
  ~~\multirow{4}{*}{22} &  ~~$\Delta\Xi_{cc}$ &          ~~5026 &    ~~5031 &   ~~ub &  ~~\multirow{4}{*}{4961} &   ~~\multirow{4}{*}{ub} \\ \cline{3-5}
  ~~  &  ~~$\Delta\Xi^{*}_{cc}$ &                  ~~5055 &   ~~5058 &    ~~ub   &    ~~ &     \\ \cline{3-5}
  ~~  &  ~~$\mathbf{\Sigma_{c}\Sigma^*_{c}}$ &   ~~4954 &   ~~4962 &    ~~ub   &    ~~ &     \\ \cline{3-5}
  ~~  &  ~~$\Sigma^*_{c}\Sigma^*_{c}$ &            ~~4983 &   ~~4990 &    ~~ub   &    ~~ &     \\ \hline
  ~~\multirow{1}{*}{23} &  ~~$\mathbf{\Delta\Xi^{*}_{cc}}$ &       ~~5055 &   ~~5063 &   ~~ub &  ~~\multirow{1}{*}{5063} &  ~~\multirow{1}{*}{ub}  \\
\hline\hline
\end{tabular}
\end{table*}

For the system with $IJ=00$, the single-channel calculation shows that both the $\Sigma_{c}\Sigma_{c}$ and $\Sigma^*_{c}\Sigma^*_{c}$ are bound states, with binding energy of $-75$ MeV and $-87$ MeV, respectively, while both $N\Xi_{cc}$ and $\Lambda_{c}\Lambda_{c}$ are unbound. This is reasonable. As shown in Fig. 1(a) that the interaction between two $\Sigma_{c}$'s (or $\Sigma^*_{c}$'s) is strong enough to form the bound state, while the attraction between $N$ and $\Xi_{cc}$ is too weak to tie the two particles together. At the same time, due to the repulsive interaction between two $\Lambda_{c}$'s, the energy of $\Lambda_{c}\Lambda_{c}$ is above its threshold. However, the effect of the channel-coupling cannot be ignored. By coupling these four channels, the lowest energy of the system is $19$ MeV lower than the threshold of $\Lambda_{c}\Lambda_{c}$, which means that the doubly-charm dibaryon system with $IJ=00$ is bound. This conclusion is consistent with the one on the hadron level~\cite{Liu2}. Besides, further work should be done to search for any resonance state. Here, we change the size of the space, which is the distance between two baryons, to see if there is any stable energy, which is corresponding to a resonance state. We find that there is a stable energy around $4716$ MeV by changing the distance from $5.0$ fm to $8.0$ fm. This energy is lower than the theoretical threshold of $N\Xi_{cc}$ about $17$ MeV. We also calculate the percentages of coupling channels for this eigen-state, and find that the proportion of $N\Xi_{cc}$ is about $41.1\%$, larger than the one of other channels, which indicates that the main component of this resonance is $N\Xi_{cc}$. This result is consistent with the conclusion of Ref.~\cite{Fromel}. However, since we do not find other stable energy by changing the space, neither the singly bound state $\Sigma_{c}\Sigma_{c}$ nor $\Sigma^*_{c}\Sigma^*_{c}$ survive as a resonance state during the channel-coupling calculation. This is understandable. Since the coupling between $\Sigma_{c}\Sigma_{c}$, $N\Xi_{cc}$, $\Lambda_{c}\Lambda_{c}$ and $\Sigma^*_{c}\Sigma^*_{c}$ is through the central force. It is strong enough to lower the energy of the $\Lambda_{c}\Lambda_{c}$ and $N\Xi_{cc}$, and meanwhile push the energy of $\Sigma_{c}\Sigma_{c}$ and $\Sigma^*_{c}\Sigma^*_{c}$ above their thresholds.

For the system with $IJ=01$, it includes three channels: $N\Xi_{cc}$, $\Sigma_{c}\Sigma^*_{c}$, and $N\Xi^*_{cc}$. The single-channel calculation shows that $\Sigma_{c}\Sigma^*_{c}$ is bound, and the binding energy is $-94$ MeV. The lowest energy of this system is $4733$ MeV after the channel-coupling calculation, which is still higher than the threshold of the lowest channel $N\Xi_{cc}$. This indicates that the doubly-charm dibaryon system with $IJ=01$ is unbound. Meanwhile, we change the distance between two baryons, and no stable energy value is found, which means that $\Sigma_{c}\Sigma^*_{c}$ maybe not a resonance state by the effect of channel-coupling.

For the system with $IJ=02$, it includes three channels: $N\Xi^*_{cc}$, $\Sigma_{c}\Sigma^*_{c}$, $\Sigma^*_{c}\Sigma^*_{c}$. The single-channel calculation shows that the energy of each single channel is higher than the corresponding threshold, indicating that none of these three channels is bound. However, the lowest energy of the system is pushed to $4741$ MeV by the channel-coupling calculation, $21$ MeV lower than the threshold of the lowest channel $N\Xi^*_{cc}$, which indicates that the doubly-charm dibaryon system with $IJ=02$ is bound. Besides, no any resonance state is found by changing the distance between two baryons here.

For the system with $IJ=10$, the result is similar to that of the $IJ=01$ system. Although the single channel $\Delta\Xi^*_{cc}$ is bound, there is no any bound state or resonance state by the channel-coupling calculation.

For the system with $IJ=11$, there are nine channels as shown in Table~\ref{result1}. The lowest energy of this system is $4739$ MeV after the channel-coupling calculation, higher than the threshold of the lowest channel $N\Xi_{cc}$, which indicates that the doubly-charm dibaryon system with $IJ=11$ is unbound. Besides, although the single-channel calculation shows that four states $\Delta\Xi_{cc}$, $\Delta\Xi^{*}_{cc}$, $\Sigma_{c}\Sigma^*_{c}$ and $\Sigma^*_{c}\Sigma^*_{c}$ are bound, there is only one resonance state survives by the effect of the channel-coupling. With the variation of the distance between two baryons, a stable energy around $4757$ MeV is obtained, which is lower than the threshold of $N\Xi^*_{cc}$. So the $N\Xi^*_{cc}$ with $IJ=11$ appears as a resonance state by the effect of the channel-coupling.

For the system with $IJ=12$, the situation is similar to that of the $IJ=11$ system. By the channel-coupling calculation, there is no any bound state, but there is a resonance state with the energy around $4949$ MeV, and the main component is $\Sigma_{c}\Sigma^*_{c}$. So the $\Sigma_{c}\Sigma^*_{c}$ with $IJ=12$ appears as a resonance state by the influence of the channel-coupling.

For the system with $IJ=13$, it includes two channels: $\Delta\Xi^*_{cc}$ and $\Sigma^*_{c}\Sigma^*_{c}$. The single-channel calculation shows that the $\Sigma^*_{c}\Sigma^*_{c}$ is a bound state with a binding energy of $-3$ MeV. After the channel-coupling calculation, the lowest energy of the system is reduced to $4969$ MeV, $14$ MeV lower than the threshold of the lower channel $\Sigma^*_{c}\Sigma^*_{c}$, which indicates that the doubly-charm dibaryon system with $IJ=13$ is bound. However, no any resonance state is found by changing the distance between two baryons here.

For the system with $IJ=20$, the result is similar to that of the $IJ=01$ system. Although the single channel $\Delta\Xi^*_{cc}$ is bound, there is no any bound state or resonance state by the channel-coupling calculation.

The results of systems with quantum numbers of $IJ=21$, $22$, and $23$ are similar. There are no bound states in either single-channel or channel-coupling calculation.
Moreover, no any resonance state is obtained either.

Because of the heavy flavor symmetry, we also extend the study to the doubly-bottom dibaryon systems. All the results are listed in Table~\ref{result2}, which are similar to those of the doubly-charm dibaryon systems. By the channel-coupling calculation, three bound systems are obtained, with the quantum numbers of $IJ=00$, $IJ=02$, and $IJ=13$, and the binding energies of $-20$ MeV, $-17$ Mev, and $-3$ Mev, respectively. Besides, three resonance states are obtained, which are $N\Xi_{bb}$ with $IJ=00$, $N\Xi^*_{bb}$ with $IJ=11$ and $\Sigma_{b}\Sigma^*_{b}$ with $IJ=12$, and the resonance mass are $11411$ MeV, $11432$ MeV and $11626$ MeV, respectively.

\begin{table*}[ht]
\caption{The energy of the doubly-bottom dibaryon systems..}   \label{result2}
\centering
\begin{tabular}{ccccccc}
\hline \hline
   ~~ $IJ$ &  ~~ $Channels$ &  ~~$E_{th}$~(MeV) &  ~~$E_{sc}$~(MeV) &  ~~$B_{sc}$~(MeV) & ~~$E_{cc}$~(MeV) &  ~~$B_{cc}$~(MeV)
  \\ \hline
  ~~\multirow{4}{*}{00} &  ~~$\Sigma_{b}\Sigma_{b}$ &    ~~11618 &   ~~11539 &   ~~-79 &  ~~\multirow{4}{*}{11219} &  ~~\multirow{4}{*}{-20}    \\ \cline{3-5}
  ~~  &  ~~$N\Xi_{bb}$ &                           ~~11424 &   ~~11428 &    ~~ub  &     ~~ & \\ \cline{3-5}
  ~~  &  ~~$\mathbf{\Lambda_{b}\Lambda_{b}}$ &  ~~11239 &   ~~11246 &    ~~ub  &     ~~ &      \\ \cline{3-5}
  ~~  &  ~~$\Sigma^*_{b}\Sigma^*_{b}$ &            ~~11636 &   ~~11553 &    ~~-83 &     ~~ &      \\ \hline
  ~~\multirow{3}{*}{01} &  ~~$\mathbf{N\Xi_{bb}}$ &    ~~11424 &   ~~11430 &   ~~ub &   ~~\multirow{3}{*}{11428} &   ~~\multirow{3}{*}{ub}   \\ \cline{3-5}
  ~~  &  ~~$\Sigma_{b}\Sigma^*_{b}$ &              ~~11627 &   ~~11544 &    ~~-83 &     ~~ &     \\ \cline{3-5}
  ~~  &  ~~$N\Xi^*_{bb}$ &                         ~~11433 &   ~~11441 &    ~~ub  &     ~~ &     \\ \hline
  ~~\multirow{3}{*}{02}&  ~~$\mathbf{N\Xi^*_{bb}}$ &   ~~11433 &   ~~11436 &   ~~ub &   ~~\multirow{3}{*}{11416} &   ~~\multirow{3}{*}{-17}  \\\cline{3-5}
  ~~  &  ~~$\Sigma_{b}\Sigma^*_{b}$ &              ~~11627 &   ~~11637 &     ~~ub &     ~~ &     \\ \cline{3-5}
  ~~  &  ~~$\Sigma^*_{b}\Sigma^*_{b}$ &            ~~11636 &   ~~11646 &     ~~ub &     ~~ &     \\ \hline
  ~~\multirow{3}{*}{10} &  ~~$\mathbf{N\Xi_{bb}}$ &    ~~11424 &   ~~11431 &   ~~ub &   ~~\multirow{3}{*}{11430} &   ~~\multirow{3}{*}{ub} \\ \cline{3-5}
  ~~  &  ~~$\Lambda_{b}\Sigma_{b}$ &               ~~11428 &   ~~11435 &     ~~ub &     ~~ &     \\ \cline{3-5}
  ~~  &  ~~$\Delta\Xi^*_{bb}$ &                    ~~11727 &   ~~11635 &     ~~-92 &    ~~ &     \\ \hline
  ~~\multirow{9}{*}{11} &  ~~$N\Xi^*_{bb}$ &             ~~11433 &   ~~11439 &   ~~ub &   ~~\multirow{9}{*}{11429} &  ~~\multirow{9}{*}{ub} \\ \cline{3-5}
  ~~  &  ~~$\Lambda_{b}\Sigma^*_{b}$ &             ~~11438 &   ~~11444 &     ~~ub &     ~~ &     \\ \cline{3-5}
  ~~  &  ~~$\Delta\Xi_{bb}$ &                      ~~11717 &   ~~11678 &     ~~-39 &    ~~ &     \\ \cline{3-5}
  ~~  &  ~~$\Delta\Xi^*_{bb}$ &                    ~~11727 &   ~~11648 &     ~~-79 &    ~~ &     \\ \cline{3-5}
  ~~  &  ~~$\Sigma_{b}\Sigma^*_{b}$ &              ~~11627 &   ~~11614 &     ~-13~ &    ~~ &     \\ \cline{3-5}
  ~~  &  ~~$\Sigma^*_{b}\Sigma^*_{b}$ &            ~~11636 &   ~~11631 &     ~~-5 &     ~~ &     \\ \cline{3-5}
  ~~  &  ~~$\Sigma_{b}\Sigma_{b}$ &                ~~11618 &   ~~11621 &     ~~ub &     ~~ &     \\ \cline{3-5}
  ~~  &  ~~$\mathbf{N\Xi_{bb}}$ &                ~~11424 &   ~~11431 &     ~~ub &     ~~ &     \\ \cline{3-5}
  ~~  &  ~~$\Lambda_{b}\Sigma_{b}$ &               ~~11428 &   ~~11435 &     ~~ub &     ~~ &     \\ \hline
  ~~\multirow{5}{*}{12} &  ~~$\Sigma_{b}\Sigma^*_{b}$ &  ~~11627 &   ~~11630 &   ~~ub &   ~~\multirow{5}{*}{11439} &  ~~\multirow{5}{*}{ub} \\ \cline{3-5}
  ~~  &  ~~$\Delta\Xi_{bb}$ &                   ~~11717 &   ~~11667 &   ~~-50 &    ~~ &     \\ \cline{3-5}
  ~~  &  ~~$\Lambda_{b}\Sigma^*_{b}$ &             ~~11438 &   ~~11444 &   ~~ub  &    ~~ &     \\ \cline{3-5}
  ~~  &  ~~$\mathbf{N\Xi^*_{bb}}$ &              ~~11433 &   ~~11441 &   ~~ub  &    ~~ &     \\ \cline{3-5}
  ~~  &  ~~$\Delta\Xi^*_{bb}$ &                    ~~11727 &   ~~11671 &   ~~-56 &    ~~ &     \\ \hline
  ~~\multirow{2}{*}{13}  &  ~~$\Delta\Xi^*_{bb}$ &       ~~11727 &   ~~11689 &   ~~-38 &    ~~\multirow{2}{*}{11633} &  ~~\multirow{2}{*}{-3} \\\cline{3-5}
  ~~  &  ~~$\mathbf{\Sigma^{*}_{b}\Sigma^{*}_{b}}$ &~~11636 &   ~~11638 &   ~~ub &    ~~ &     \\ \hline
  ~~\multirow{3}{*}{20} &  ~~$\Delta\Xi^*_{bb}$ &        ~~11727 &   ~~11721 &   ~~-6 &   ~~\multirow{3}{*}{11621} &   ~~\multirow{3}{*}{ub} \\ \cline{3-5}
  ~~  &  ~~$\Sigma^*_{b}\Sigma^*_{b}$ &            ~~11636 &   ~~11640 &   ~~ub &     ~~ &     \\ \cline{3-5}
  ~~  &  ~~$\mathbf{\Sigma_{b}\Sigma_{b}}$ &     ~~11618 &   ~~11622 &   ~~ub &     ~~ &     \\ \hline
  ~~\multirow{3}{*}{21} &  ~~$\textbf{$\Sigma_{b}\Sigma^*_{b}$}$ &  ~~11627 &   ~~11632 &   ~~ub  &~~\multirow{3}{*}{11631} &   ~~\multirow{3}{*}{ub} \\ \cline{3-5}
  ~~  &  ~~$\Delta\Xi_{bb}$ &                      ~~11717 &   ~~11724 &   ~~ub &     ~~ &     \\ \cline{3-5}
  ~~  &  ~~$\Delta\Xi^*_{bb}$ &                    ~~11727 &   ~~11726 &   ~~-1 &     ~~ &     \\ \hline
  ~~\multirow{4}{*}{22} &  ~~$\Delta\Xi_{bb}$ &          ~~11717 &   ~~11722 &   ~~ub &  ~~\multirow{4}{*}{11633} &   ~~\multirow{4}{*}{ub} \\ \cline{3-5}
  ~~  &  ~~$\Delta\Xi^{*}_{bb}$ &                   ~~11727 &   ~~11731 &  ~~ub &     ~~ &     \\ \cline{3-5}
  ~~  &  ~~$\mathbf{\Sigma_{b}\Sigma^*_{b}}$ &   ~~11627 &   ~~11634 &   ~~ub &     ~~ &     \\ \cline{3-5}
  ~~  &  ~~$\Sigma^*_{b}\Sigma^*_{b}$ &            ~~11636 &   ~~11643 &   ~~ub &     ~~ &     \\ \hline
  ~~\multirow{1}{*}{23} &  ~~$\mathbf{\Delta\Xi^{*}_{bb}}$ &    ~~11727 &   ~~11734 &   ~~ub &  ~~\multirow{1}{*}{11734} &  ~~\multirow{1}{*}{ub}  \\
\hline\hline
\end{tabular}
\end{table*}

\section{Summary}
The low-lying doubly-heavy dibaryon systems with strange $S=0$, isospin $I=0$, $1$, $2$ and the angular momentum $J=0$, $1$, $2$, $3$ are systemically investigated by using the RGM in the framework of QDCSM. Our goal is to search for any bound state or resonance state of the doubly-heavy dibaryon systems. The effective potential of every channel is calculated to observe the interaction between two baryons. Both single-channel and channel-coupling calculations are performed to obtain the energy of all the systems. Besides, to search for any resonance state, a calculation of changing the distance between two baryons is carried out.

The numerical results show that for the doubly-charm dibaryon systems, there are three bound systems, the quantum numbers of which are $IJ=00$, $IJ=02$ and $IJ=13$, and the energies are $4554$ MeV, $4741$ MeV, and $4969$ MeV, respectively. Besides, three resonance states $N\Xi_{cc}$, $N\Xi^*_{cc}$ and $\Sigma_{c}\Sigma^*_{c}$ are obtained with the quantum numbers of $IJ=00$, $IJ=11$ and $IJ=12$, and the resonance mass of $4716$ MeV, $4757$ MeV, $4949$ MeV, respectively.
Similarly, for the doubly-bottom dibaryon systems, the quantum numbers of three bound systems are $IJ=00$, $IJ=02$ and $IJ=13$, and the energies are $11219$ MeV, $11416$ MeV, and $11633$ MeV, respectively. Additionally, three resonance states are $N\Xi_{bb}$ with $IJ=00$, $N\Xi^*_{bb}$ with $IJ=11$ and $\Sigma_{b}\Sigma^*_{b}$ with $IJ=12$, and the resonance mass are $11411$ MeV, $11432$ MeV, $11626$ MeV, respectively. All these heavy dibaryons are worth looking for on experiments, although it will be a challenging subject.

We also find that the effect of the channel-coupling is remarkable in the study of multi-quark systems. In this work, some single states are not bound at first, but they become bound by the channel-coupling calculation. Meanwhile, some states, which are bound in the single channel calculation, appear as unbound states or resonance states after the channel-coupling calculation. The main reason is that the doubly-heavy dibaryon systems we investigate here is all in $S-$ wave, and the channel-coupling between all channels is through the central force, the role of which has been verified to be much more important than the tensor force in our quark level calculation~\cite{Huang_LL}. Therefore, to explore the multi-quark states, the effect of channel-coupling cannot be neglected. Besides, we search for the resonance states only by changing the size of the orbital space. The study of the scattering process of the corresponding open channels are needed to confirm the existence of resonance states, which is our further work.

\acknowledgments{This work is supported partly by the National Science Foundation
of China under Contract Nos. 11675080, 11775118 and 11535005.}


\begin{thebibliography}{99}
\bibitem{ABC1} M. Bashkanov {\em et al} (CELSIUS-WASA Collaboration),
Phys. Rev. Lett. {\bf 102}, 052301 (2009).
\bibitem{ABC2} P. Adlarson {\em et al} (WASA-at-COSY Collaboration),
Phys. Rev. Lett. {\bf 106}, 242302 (2011).
\bibitem{ABC3} P. Adlarson {\em et al} (WASA-at-COSY Collaboration),
Phys. Lett. {\bf B721}, 229 (2013).
\bibitem{ABC4} P. Adlarson {\em et al} (WASA-at-COSY Collaboration),
Phys. Rev. {\bf C88}, 055208 (2013).
\bibitem{ABC5} P. Adlarson {\em et al} (WASA-at-COSY Collaboration),
Phys. Rev. Lett. {\bf 112}, 202301 (2014).
\bibitem{Dyson} F. J. Dyson and N. H. Xuong, Phys. Rev. Lett. {\bf 13}, 815 (1964).
\bibitem{Goldman} T. Goldman, K. Maltman, G. J. Stephenson, K. E. Schmidt and F. Wang,
 Phys. Rev. {\bf C39}, 1889 (1989).
\bibitem{faddeev} A. Gal and H Garcilazo, Phys. Rev. Lett. {\bf 111}, 172301 (2013).
\bibitem{Bashkanov} M. Bashkanov, S. J. Brodsky and H. Clement, Phys. Lett. B {\bf 727}, 438
(2013).
\bibitem{Ping_NN} J. L. Ping, H. X. Huang, H. R. Pang, F. Wang and C. W. Wong, Phys. Rev. {\bf C 79}, 024001
(2009).
\bibitem{Huang_NN} H. X. Huang, J. L. Ping and F. Wang, Phys. Rev. {\bf C 89}, 034001
(2014).
\bibitem{DongYB_NN} Y. Dong, P. Shen, F. Huang and Z. Zhang, Phys. Rev. {\bf C 91}, 064002
(2015).
\bibitem{ChenHX_NN} H. X. Chen, E. L. Cui, W. Chen, T. G. Steele and S. L. Zhu, Phys. Rev. {\bf C 91}, 025204
(2015).
\bibitem{PRL59} T. Goldman, K. Maltman, G. J. Stephenson, K. E. Schmidt and F. Wang,
Phys. Rev. Lett. {\bf 59}, 627 (1987).
\bibitem{Oka} M. Oka, Phys. Rev. D {\bf 38}, 298 (1988).
\bibitem{PRC51} F. Wang, J. L. Ping, G. H. Wu, L. J. Teng and T. Goldman,  Phys. Rev. C {\bf 51}, 3411 (1995).
\bibitem{PangHR} H. R. Pang, J. L. Ping, F. Wang, T. Goldman and E. G. Zhao, Phys. Rev. C {\bf 69}, 065207 (2004).
\bibitem{ChenM} M. Chen, H. X. Huang, J. L. Ping and F. Wang, Phys. Rev. C {\bf 83}, 015202 (2011).
\bibitem{Huang2} H. X. Huang, J. L. Ping and F. Wang, Phys. Rev. C {\bf 92}, 065202 (2015).
\bibitem{LiQB} Q. B. Li, P. N. Shen, Eur. Phys. J. A {\bf 8}, 417 (2000).
\bibitem{HAL1} F. Etminan, {\em et al.} (HAL QCD Collaboration), Nucl. Phys. A {\bf 928}, 89 (2014).
\bibitem{HAL2} T. Iritani, {\em et al.} (HAL QCD Collaboration), Phys. Lett. B {\bf 792}, 284 (2019).
\bibitem{RHIC} J. Adam {\em et al} (STAR Collaboration), Phys. Lett. B {\bf 790}, 490
(2019).
\bibitem{ALICE} S. Acharya {\em et al} (ALICE Collaboration), Nature {\bf 588}, 236
(2020).
\bibitem{Liu1} Y. R. Liu and M. Oka, Phys. Rev. D {\bf 85}, 014015 (2012).
\bibitem{Liu2} W. Meguro, Y. R. Liu and M. Oka, Phys. Lett. B {\bf 704}, 547 (2011).
\bibitem{Huang_NL} H. X. Huang, J. L. Ping and F. Wang, Phys. Rev. C {\bf 87}, 034002 (2013).
\bibitem{Huang_LL} H. X. Huang, J. L. Ping and F. Wang, Phys. Rev. C {\bf 89}, 035201 (2014).
\bibitem{Fromel} F. Fromel, B. Julia-Diaz and D. O. Riska, Nucl. Phys. A {\bf 750}, 337 (2005).
\bibitem{Julia} B. Julia-Diaz and D. O. Riska, Nucl. Phys. A {\bf 755}, 431 (2005).
\bibitem{MengL} Lu Meng, Ning Li and Shi-Lin Zhu (2017), arXiv:1704.01009v1.
\bibitem{Huang_NO} H. X. Huang and J. L. Ping, Phys. Rev. C {\bf 101}, 015204 (2020).
\bibitem{Lattice_De} P. Junnarkar and N. Mathur, Phys. Rev. Lett. {\bf 123}, 162003 (2019).
\bibitem{Richard} J. M. Richard, A. Valcarce, and J. Vijande, Phys. Rev. Lett. {\bf 124}, 212001 (2020).
\bibitem{Huang_OO} H. X. Huang, J. L. Ping, X. M. Zhu and F. Wang, arXiv: 2011.00513.
\bibitem{QDCSM0} F. Wang, G. H. Wu, L. J. Teng and T. Goldman, Phys. Rev. Lett. {\bf 69}, 2901 (1992);
G. H. Wu, L. J. Teng, J. L. Ping, F. Wang and T. Goldman, Phys. Rev. C {\bf 53}, 1161 (1996).
\bibitem{QDCSM1} J. L. Ping, F. Wang and T. Goldman, Nucl. Phys. A {\bf 657}, 95 (1999); G. H. Wu, J. L. Ping,
L. J. Teng {\em et al.}, Nucl. Phys. A {\bf 673}, 279 (2000); H. R. Pang, J. L. Ping, F. Wang and T. Goldman, Phys. Rev. C {\bf 65}, 014003 (2001); J. L. Ping, F. Wang and T. Goldman, Nucl. Phys. A {\bf 688}, 871 (2001); J. L. Ping, H. R. Pang, F. Wang and T. Goldman, Phys. Rev. C {\bf 65}, 044003 (2002).
\bibitem{QDCSM2} J. L. Ping, H. X. Huang, H. R. Pang, F. Wang and C. W. Wong, Phys. Rev. C {\bf 79}, 024001 (2009);
\bibitem{QDCSM3} M. Chen, H. X. Huang, J. L. Ping and F. Wang, Phys. Rev. C {\bf 83}, 015202 (2011).
\bibitem{QDCSM4} L. Z. Chen, H. R. Pang, H. X. Huang, J. L. Ping and F. Wang, Phys. Rev. C {\bf 76}, 014001 (2007);
\bibitem{QDCSM5} H. X. Huang, P. Xu, J. L. Ping and F. Wang, Phys. Rev. C {\bf 84}, 064001 (2011).
\bibitem{Salamanca} A. Valcarce, H. Garcilazo, F. Fern\'{a}ndez and P. Gonzalez, Rep. Prog. Phys. {\bf 68}, 965 (2005) and references therein.
\bibitem{JPG31} J. Vijande, F. Fernandez and A. Valcarce, J. Phys. G {\bf 31}, 481 (2005).
\bibitem{PDG} C. Patrignani, {\em et al.}, Particle Data Group, Chinese Phys. C {\bf 40}, 100001 (2016).
\bibitem{Xu} M. M. Xu, M. Yu and L. S. Liu, Phys. Rev. Lett. {\bf 100},
  092301 (2008).
\bibitem{RGM} M. Kamimura, Supp. Prog. Theo. Phys. {\bf 62}, 236 (1977).
\end{thebibliography}
\end{document}